\documentclass[fleqn,usenatbib]{mnras}
\usepackage{newtxtext,newtxmath}
\usepackage[section]{placeins}
\usepackage{multirow}
\usepackage{booktabs}
\usepackage{siunitx}
\usepackage{float}
\usepackage{amsmath}
\usepackage{float}
\usepackage{appendix}
\usepackage{threeparttable}
\usepackage{placeins}

\usepackage[T1]{fontenc}

\DeclareRobustCommand{\VAN}[3]{#2}
\let\VANthebibliography\thebibliography
\def\thebibliography{\DeclareRobustCommand{\VAN}[3]{##3}\VANthebibliography}

\usepackage{graphicx}	
\usepackage{amsmath}	

\newcommand{\rion}[2]{{\ensuremath{\mbox{\rm #1$\,${\small\uppercase\expandafter{\romannumeral#2\relax}}}}}}
\newcommand{\HI}{{\rion{H}{1}}}
\newcommand{\HII}{\rion{H}{2}}

\newcommand{\OIII}{[\rion{O}{3}]}
\newcommand{\OII}{[\rion{O}{2}]}
\newcommand{\MV}{\ensuremath{\mbox{\rm \texttt{MAPPINGS\,V}}}}
\newcommand{\rss}{\color{red}}
\title[High-redshifts LF and flux ratios]{A model for the emission line galaxy luminosity function and flux ratios at high-redshifts}

\author[Pathak et al.]{
Aadarsh Pathak,$^{1,3}$\thanks{E-mail: aadarsh.pathak@student.unimelb.edu.au}
J.Stuart B. Wyithe$^*$,$^{2,3}$
Ralph S. Sutherland,$^{2}$
L.J Kewley$^{4}$
\\
\textsuperscript{\rm 1}School of Physics, University of Melbourne, Parkville, VIC 3010, Australia\\
\textsuperscript{\rm 2}Research school for Astronomy and Astrophysics, Mount Stromlo Observatory, Cotter Road, Weston, ACT 2611\\
\textsuperscript{\rm 3}ARC Centre of Excellence for All Sky Astrophysics in 3 Dimensions (ASTRO 3D)\\
\textsuperscript{\rm 4}Center for Astrophysics--Harvard and Smithsonian, 60 Garden Street, Cambridge, MA 02138, USA
}

\date{Accepted XXX. Received YYY; in original form ZZZ}

\pubyear{2015}

\begin{document}
\label{firstpage}
\pagerange{\pageref{firstpage}--\pageref{lastpage}}
\maketitle

\begin{abstract}
We present $\OIII/H_{\rm \beta}$ emission line flux ratio  predictions for galaxies at $z \sim 7-9$ using the $\MV$ v5.2.0 photoionization modelling code combined with an analytic galaxy formation model. Properties such as pressure and ionization parameter that determine emission line properties are thought to evolve towards high redshift. In order to determine the range of expected interstellar conditions we extend previous modelling of the Star Formation Rate Density (SFRD) function to calculate the metallicity and ionization parameter, and incorporate the potential impact of turbulence on the density of the ISM. To validate our emission line predictions we calculate the $\OIII$ line luminosity and its dependence on UV luminosity, as well as the flux ratio $\OIII/H_{\rm \beta}$ and its variation with the line luminosity, finding that both reproduce recent JWST observations from the FRESCO survey. We also use our model to predict the number counts of emission line galaxies across a range of redshift as well as the dependence of $\OIII/H_{\rm \beta}$ on ionization parameter and metallicity. Finally, we show that the dependence of flux ratio on luminosity may provide a diagnostic of turbulent motion in galactic discs.

\FloatBarrier
\end{abstract}



\section{Introduction}

The characterization of the physical state of the interstellar medium (ISM) is important for unraveling the mechanisms governing galaxy formation. One diagnostic for assessing the ISM state is provided by rest--frame optical emission lines, including hydrogen recombination lines and collisionally excited metal lines. These emission lines provide {\rss } insight into the local environment where stars are born. For example the ionization parameter ($q$) which measures the hydrogen ionizing photon flux with respect to the number density of hydrogen atoms serves as a key indicator of star formation distribution and establishes a connection between ionizing sources and ionized gas \citep{2012Yeh,2018kassinen,2019Lisa}. However, properties of emission lines in star-forming regions within the local universe differ significantly from those observed in high-redshift environments \citep{2013aKewley,2014Steidel,Shapley_2015}.

Numerous studies have documented an elevation in the ionization parameter at higher redshifts \citep{2015lisa,2016Sanders,2018Kewley}, which correlates with 
increased ${\OIII}~\lambda \lambda 4959,5007/{\OII}~\lambda \lambda 3726,3729$, and ${\OIII}~\lambda 5007 / H_{\rm \beta}$ ratios when compared to local galaxies. This observation signifies a shift in the ionization state of the ISM and has implications for { understanding} evolving galactic environments. Another crucial indicator of the ISMs physical condition is the gas-phase metallicity. In the local universe, there exists a positive correlation between gas-phase metallicity and stellar mass ($M_*$), as defined by the Mass-Metallicity Relation (MZR) \citep{2004Tremonti,2006Lee,2008Kewley,2013Andrews,2019Blanc}. However, when holding stellar mass ($M_*$) constant, the MZR undergoes evolution where gas-phase metallicity decreases with increasing redshift \citep{2005Savaglio,2009Mannucci,2011Zahid,2014Zahida,2014Zahidb,2014Cullen,2014Steidel,2016Hunt,2017Suzuki}. The evolution of the MZR has also been explored through theoretical simulations. In particular, the IllustrisTNG simulation has examined its redshift evolution and scatter, offering insights into the underlying physical processes driving metallicity variations across cosmic time \citep{2019Torrey}. Recent spectroscopic observations with JWST have significantly advanced the study of the MZR relation, extending its measurement to very high redshifts, up to $z \sim 10$ \citep{2024Chemerynska, Nakajima2023, Curti2023}.

The James Webb Space Telescope (JWST) has provided the opportunity to spectroscopically study Epoch of Reionization galaxies for the first time. For example, the JWST FRESCO survey, operating at wavelengths of approximately $4-5$ micrometers ($\mu$m), has assembled a {significant} sample of star-forming emission line galaxies \citep{2023FRESCO} down to luminosities of approximately $0.2-0.5$ times that of the characteristic luminosity $(L_*)$ in the rest-frame ultraviolet band. Emission line properties have also been studied using the Cosmic Evolution Early Release Science survey, CEERS \citep{Finkelstein_2023}, EIGER survey \citep{kashino2023, 2023bMatthee} and JWST Advanced Deep Extragalactic Survey (JADES) \citep{2023Cameron}. In this study, we introduce a model to predict strong emission line ratios for high-redshift galaxies. To achieve this, we utilize the Star Formation Rate Density (SFRD) model discussed previously in \cite{Stu2012}. The model is based on the extended Press-Schechter formalism \citep{PSMF1974,1999sheth} assuming that each major merger triggers star formation lasting for a time $ t_{\rm SF}$ and converts at most $f_{\rm *,max}$ of the galactic gas into stars. This model incorporates the effect of supernova (SN) feedback in regulating star formation. The model explains the amplitude and shape of the SFRD between redshifts 4 and 7. This model can be used to calculate several key ISM properties, including pressure, particle number density, ionization parameter, and metallicity. We extend this model to explore how turbulence impacts various physical properties and estimate the line luminosity and luminosity function of the rest-frame optical emission line {$\OIII \lambda 5007$.} Beyond observational constraints, several hydrodynamical simulation studies investigated the flux ratios and luminosity functions (LFs). For instance, \cite{2023Hirschmann} employ the IllustrisTNG cosmological simulation, incorporating a self-consistent photoionization model to compute flux ratios and LFs including fast radiative shocks, young star clusters, active galactic nuclei (AGN), and post-AGB stellar populations to derive emission line properties. They reported an increase in $\OIII/H_{\rm \beta}$ at fixed stellar mass from low to high redshift, primarily driven by a higher ionization parameter in high-redshift galaxies, regulated by an elevated specific star formation rate and a greater global gas density. On the other hand, the SPHINX public data release \citep{2023katz} offers significantly higher spatial and mass resolution compared to IllustrisTNG. This leads to a more bursty star formation history, introducing greater scatter in the UV magnitude and stellar mass ($M_{\rm *}$). They found $\OIII/H_{\rm \beta}>10$ for many galaxies attributed to their ionization parameter being significantly greater than -2. Complimenting these studies, our analytic model is used to efficiently calculate ISM conditions with detailed photoionization calculations of emission line properties. We calculate the $\OIII/H_{\rm \beta}$ flux ratio using the $\MV$ v5.2.1 \footnote{Suttherland Priv Comm: \url{https://mappings.anu.edu.au} for the public $\MV$ v5.2.1 version 2023.} photoionization modelling code and show its variation with {$\OIII \lambda 5007$} line luminosity.

The structure of this paper is as follows: In section \ref{sec:SFRD}, we discuss the SFRD model, ISM turbulence and various physical properties. In section \ref{sec:results}, we describe the effect of turbulence on physical properties. Section \ref{sec:emission_line} outlines the line luminosity calculation and flux ratio estimation. In section \ref{sec:LF}, we focus on {the} luminosity function and number counts for FRESCO emission line galaxies, and in section \ref{sec:line_ratio} we describe the $\OIII/H_{\rm \beta}$ line ratio calculation. In section \ref{sec:cav}, we discuss the caveats of the model. Finally in section \ref{sec:summary}, we present our summary and conclusion.

\section{Methods}
\label{sec:SFRD}
\subsection{Analytic galaxy formation model}
We start with a description of the analytic Star Formation Rate Density function (SFRD) model from \cite{Stu2012} that incorporates the role of major mergers in initiating star formation and the regulatory effects of Supernovae (SNe) feedback. In this model, the Star Formation Rate Density (galaxies per {\rm M\,pc$^{-3}$} per unit SFR) function\footnote{In order to compare the model with observation, one can rewrite \ref{eq:SFRD} is \\
    $\psi(SFR) = \ln 10 \times SFR \times \phi$ \\
    which has units of {\rm M\,pc$^{-3}$ per dex.}} is 
\begin{equation}
    \phi(SFR) = \epsilon_{\rm duty} \ \frac{dn}{dM_{\rm halo}} \ \left(\frac{dSFR}{dM_{\rm halo}} \right)^{-1} \;,
    \label{eq:SFRD}
\end{equation}

where, $\epsilon_{\rm duty}$ is the fraction of Hubble time for which a galaxy is visible as a starburst and $dn/dM_{\rm halo}$ is the halo mass function \citep{PSMF1974,1999sheth}.
The SFR is 
\begin{equation}
    SFR = \frac{\Omega_{\rm b}}{\Omega_{\rm m}} (\rm M_{\odot}/yr) \  \left(\frac{m_{\rm d}}{0.17}\right) \left(\frac{M_{\rm halo}}{10^8 M_{\odot}}\right) \ \left(\frac{f_{*}}{0.035}\right) \ \left( t_{\rm SF} \right),
    \label{eq:SFR}
\end{equation}

where $(m_{\rm d}M_{\rm halo})$ is the disc mass and $f_{*}$ is the star formation efficiency with which the disc mass is being converted into stars over a time $t_{\rm SF}$ (i.e, the starburst duration).\\
The duty cycle ($\epsilon_{\rm duty}$) in equation (\ref{eq:SFRD}) is given by
\begin{equation}
     \epsilon_{\rm duty} = \left(\frac{t_{\rm s}+t_{\rm SF}}{t_{\rm H}}\right)N_{\rm mergers} \; ,
    \label{eq:duty}
\end{equation}
where {$t_{\rm s} \sim 3 - 4  \times 10^6 $\,\rm years} is the time-scale over which the most massive stars {evolve and die} away. The model assumes each major merger triggers star formation. The rate of mergers $(dN_{\rm mergers}/dt)$ is calculated based on the work of \citet{Lacey1994} as the number of haloes per logarithm of mass $\Delta M$ per unit time merge with another halo of mass $M_1$ to form a halo of mass {$M_{\rm halo}$} (see \citealp{Stu2012} for details).\\

The {gas phase particle} number density, $n_{\rm p}$, is needed to evaluate the ionization state of the disc. In the mid plane of the disc at the scale radius, $R_{\rm d}$, we evaluate $n_{\rm p}$ from hydrostatic equilibrium
{\begin{equation}
    n_{\rm p} \approx \frac{G \ (m_{\rm d} \ M_{\rm halo})^{2}}{8 \ \pi \ m_{\rm p} \ c_{\rm s}^2 \ R_{\rm d}^4 \times 2.71^2} \; ,
    \label{eq:npd}
\end{equation}}
where {$G$} is the universal gravitational constant, $m_{\rm p}$ is the mass of {the} proton, $c_{\rm s}$ is the adiabatic speed of sound. The disc radius $R_{\rm d}$ is calculated assuming adiabatic contraction and is given by
\begin{equation}
    R_{\rm d} = \frac{\lambda}{\sqrt{2}}  \ R_{\rm vir}, \;
    \label{eq:rscale}
\end{equation}
where, $\lambda$ is spin parameter which we {take} as 0.05 \citep{1998Mo} and $R_{\rm vir}$ is the virial radius of the disc defined as
\begin{equation}
    R_{\rm vir} = 0.784 \ {\rm h^{-1} kpc} \ \left(\frac{M_{\rm halo}}{10^8 \rm M_{\odot} h}\right)^{\frac{1}{3}} \ [\zeta(z)]^{-\frac{1}{3}} \ \left(\frac{1+z}{10}\right)^{-1} \; ,
    \label{eq:R_vir}
\end{equation}\\
with $\zeta(z) = [(\Omega_{\rm m}/\Omega_{\rm m}^z)(\Delta_{\rm c}/18 \pi^2]$ and $\Omega_{\rm m}^z = [1+(\Omega_{\lambda}/\Omega_{\rm m})(1+z)^{-3}]^{-1}$, $\Delta_{\rm c} = 18\pi^2 + 82d - 39d^2$  and $d=\Omega_{\rm m}^z -1$ \citep{Barkana2001}.

\subsubsection{{Supernovae} feedback on star formation}

The calculation implements the galactic porosity model originally proposed by \citet{Clarke2002} to incorporate the effect of SNe feedback on the ISM. It assumes that clusters of $N_{\rm SN}$ SNe produce superbubbles in the ISM having a radius $R_{\rm e}$ which is

\begin{equation}
    \begin{split}
        R_{\rm e} = 0.08 \ {\rm kpc} \ \left(\frac{N_{\rm SN}}{10}\right)^{1/3}\ \left(\frac{E_{\rm SN}}{10^{51}\rm erg}\right)^{1/3} \ \left(\frac{\lambda}{0.05}\right)^{4/3} \\ 
        \times \left(\frac{m_{\rm d}}{0.17}\right)^{-2/3} \ \left(\frac{M_{\rm halo}}{10^8 \ \rm M_\odot}\right)^{-2/9} \ \left(\frac{1+z}{10}\right)^{-4/3} \; .
        \label{eq:R_e}
    \end{split}    
    \end{equation}\\
    
Here, $E_{\rm SN}$ is SNe energy output, {$N_{\rm SN}$ } is number of SNe in each cluster and $c_{\rm s}$ is the sound speed, assumed to be 10\,km/s {for typical $10^4$\,K gas}.    
This superbubble undergoes evacuation within the interstellar medium (ISM) on a timescale of $t_{\rm e} = 4 \times 10^7$ years. The deposit of supernova energy depends on the relationship between the superbubble's radius and the scale height of the galactic disc, denoted as {$H$}, where
\begin{equation}
\begin{split}
     H = 0.034 \, {\rm kpc} \,  \left(\frac{\lambda}{0.05}\right)^{2}
         \left(\frac{m_{\rm d}}{0.17}\right)^{-1} \ \left(\frac{M_{\rm halo}}{10^8 \ \rm M_\odot}\right)^{-1/3} \\ \times \left(\frac{1+z}{10}\right)^{-2} \ \left(\frac{c_{\rm s}}{10 \,{\rm km/s}} \right)^2 \; .
         \label{eq:H}
\end{split}         
\end{equation}

When the superbubble's radius {\rss}$R_{\rm e}$, exceeds {$H$}, only a proportion $f_{\rm d} = 2H/R_{\rm e}$ of the SN energy contributes to driving the porosity of the ISM within galactic discs. The ratio $f_{\rm d}$ is given as

\begin{equation}
\begin{split}
    f_{\rm d} = 0.85 \ \left(\frac{N_{\rm SN}}{10}\right)^{-\frac{1}{3}} \left(\frac{E_{\rm SN}}{10^{51} \,{\rm erg}}\right)^{-\frac{1}{3}}  \left( \frac{\lambda}{0.05}\right)^{\frac{2}{3}}  \left(\frac{m_{\rm d}}{0.17}\right)^{-\frac{1}{3}} \\ 
    \times \left( \frac{M_{\rm halo}}{10^8 M_{\odot}}\right)^{-\frac{1}{9}}  \left( \frac{1+z}{10}\right)^{-\frac{2}{3}} \left( \frac{c_{\rm s}}{10\,{\rm km/s}}\right)^2 \;.
    \label{eq:fd}
\end{split}    
\end{equation}
Equation \ref{eq:fd} holds until $f_{\rm d}<1$ otherwise $ f_{\rm d}=1$.\\

With these quantities, we can calculate the star formation efficiency through balancing SNe energy and disk binding energy. From \cite{2014Stu}, this is
\begin{equation}
\begin{split}
    f_* = min \left[ f_{*,\rm max},\frac{0.008}{N_{\rm merge}} \left( \frac{M_{\rm halo}}{10^{10} M_{\odot}}\right)^{\frac{2}{3}} \left( \frac{1+z}{10} \right)(f_{\rm t} f_{\rm d} F_{\rm SN})^{-1} \right ],
\end{split}    
\end{equation}
where $f_{\rm *,max}$ is the maximum star formation efficiency linked to individual major mergers, $f_{\rm d}$ is the parameter mentioned in equation \ref{eq:fd}, and $f_{\rm t}$ denotes the fraction of SN energy that contributes because of the finite timescale of SN feedback 
\begin{equation}
    f_{\rm t} =\left( \frac{t_{\rm SF}}{t_{\rm e}}\right)^2 \; .
\end{equation}
Here $t_{\rm e}$ is the timescale associated with the {superbubble  evacuation in the ISM by SNe in the cluster}. A value $t_{\rm SF}<t_{\rm e}$, signifies that not all of the SNe {\rss } energy will be accessible for {feedback processes}. We define {a critical Star Formation Rate $(SFR_{\rm crit})$ }  required to attain a porosity of unity as
\begin{equation}
    SFR_{\rm crit} = 0.15  \ {\rm M_{\odot} yr^{-1}} \ \left(\frac{m_{\rm d} M_{\rm halo}}{10^{10} \ M_{\odot}}\right) \ \left(\frac{c_{\rm s}}{10\,{\rm km/s}}\right)^2 \ (f_{\rm t} f_{\rm d})^{-1} \; .
\end{equation}
Above this critical SFR, the escape fraction for ionizing photons will reach unity.

\begin{figure}
    \centering
    \includegraphics[scale=0.45]{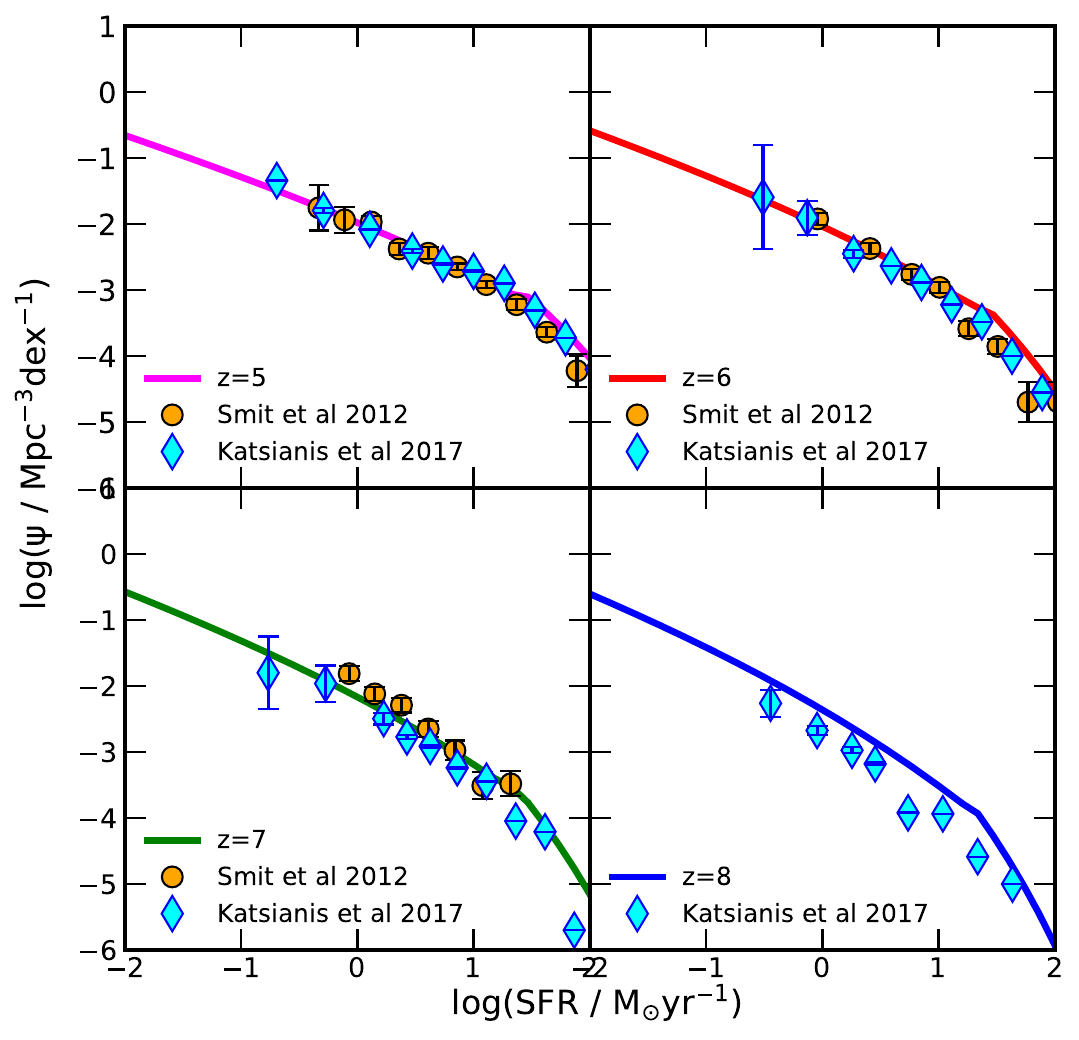}\par
    
    \caption{The comparison between the modelled and observed SFRD functions plotted for redshifts $z \sim 5,6,7,8$. The two free parameters $t_{SF}$ and $f_{\rm *,max}$ are chosen as $2.0 \times 10^7$ years and 0.035 respectively. The observed values shown by cyan and orange colored data-points correspond to the results of \citep{2017Katsianis} and \citep{Smit2012} respectively.}
    \label{fig:SFRD}
\end{figure}

In Figure \ref{fig:SFRD}, we illustrate the model SFRD plotted against the Star Formation Rate (SFR) for four redshifts: $z$ = 5, 6, 7 and 8. The modeled SFRD function is overlaid on observed SFR functions from \cite{Smit2012} shown by orange colored circles and \cite{2017Katsianis} shown by cyan colored diamonds. We set two free parameters to describe the observed Star Formation Rate Density (SFRD) function for high-redshift galaxies, finding $t_{SF} = 2.0 \times 10^7 $ years and $f_{*,max} = 0.035$, as {discussed in} \cite{Stu2012}. In the following sections, we introduce our model extensions, to derive the underlying physical characteristics of the galactic disc.

 \subsection{Effect of turbulent ISM on disc properties }
Turbulent ISM motions in the galactic disc can modify the disc scale height. ISM turbulence can result from energy sources including stellar feedback, magnetic fields, stellar winds, radial mass transport and gas accretion, \citep{2018Krumholz}. The turbulence is assumed to be sustained via an energy balance between these energy sources and its dissipation in the interstellar medium. In our analysis, we consider two main sources, supernovae feedback  and external gas accretion \citep{2016Shober,2022Ginzburg}.

\subsubsection{Turbulence driven by {supernova} feedback}
We implement the approach of \cite{2016Shober} which links the injection of turbulent kinetic energy with the SFR and the scale height of the disc{, $H$,} which sets the timescale for the turbulence decay. It assumes that in a steady state, one can find the turbulent velocity $v_{\rm SN}$ by considering the equipartition between the loss or dissipation rate,

\begin{equation}
L_{\rm loss} = \left(\frac{\rho v_{\rm SN}^2}{2 H/v_{\rm SN}} \right) \; ,
\end{equation}
and the SN energy injection rate :
\begin{equation}
L_{\rm SN} = \left(\dot{\rho}_{\rm SN} f_{\rm SN} E_{\rm SN}\right) \;
\end{equation}
On equating these two equations, one can write $v_{\rm SN}$ as 

\begin{equation}
    v_{\rm SN} = (2 \dot{\rho}_{\rm SN} f_{\rm SN} E_{\rm SN} H \rho^{-1})^{1/3},
    \label{eq:v_turb}
\end{equation}\\
where $f_{\rm SN}E_{\rm SN}$ is the fraction of SNe energy converted into turbulent energy, { $H$ is the scale height of the disc and $\rho$ is the mass density of particles given as $\rho = \mu m_{\rm p} n_{\rm p}$, where $\mu$ is the usual mean molecular weight, $m_{\rm p}$ is the proton mass and $n_{\rm p}$ is the total particle number density. The SN rate density $\dot{\rho}_{\rm SN}$ evaluated as

{ 
\begin{equation}
    \dot{\rho}_{\rm SN} = 0.156 \, \left[ \frac{SFR}{\bar{M}_{\rm SN} V} \right] \; .
    \label{eq:SNdensity}
\end{equation}
}
Here, the number density of SN per unit time is calculated by assuming a Kroupa stellar initial mass function see \citep{2016Shober} which closely approximates the common Salpeter IMF, for stars massive enough to produce SNe, ($\bar{M}_{\rm SN} \approx 12.26 \rm M_\odot$) and $\rm V$ is the disc volume. 

\subsubsection{Turbulence driven by gas accretion}
Accretion can generate turbulence through different mechanisms {including the} redistribution of angular momentum carried by accreting material through shear forces, viscosity and via disc instabilities caused by thermal processes and gravitational interactions. We consider the turbulence in our model based on the approach proposed by \citep{2022Ginzburg}. This model self-consistently considers the effect of mass-transport by balancing energy provided by mass-transport with the energy dissipation through turbulence. In the model, the halo accretion rate estimation is based on the Extended Press-Schechter (EPS) theory 
\begin{equation}
    \frac{\dot{M_{\rm halo}}}{M_{\rm halo}} = -a  M_{\rm halo,12}^b  \dot{\omega}
\end{equation}
where {$b = 0.14$, $a = 0.628$ } and $\omega$ is the self--similar time variable which can be written in a derivative form as 
\begin{equation}
    \dot{\omega} = -0.0476(1+z + 0.093 (1+z)^{-1.22})^{2.5} \ \rm Gyr^{-1} \; .
\end{equation}
Evaluating the accreted baryons as $f_{\rm b} \dot{M}_{\rm halo}$ with $f_{\rm b} \approx 0.17$, {we} can define the baryon accretion rate as 
\begin{equation}
    \dot{M}_{\rm g,acc} = \epsilon f_{\rm b} \dot{M}_{\rm halo} \; ,
\end{equation}
where $\epsilon$ is a penetration parameter to account for the efficiency with which the baryons penetrate the host halo of the galaxy, parameterized as $ min(\epsilon_0 M_{\rm halo,12}^{\rm \alpha_1},1) $. Here, the best-fit parameters for $z \geq 2$ are $\epsilon_0 = 0.31$, $\alpha_1 = -0.25$ and $\alpha_2 = 0.38$ \citep{2022Ginzburg}.

The velocity dispersion for turbulence based on the energy balance between accretion, mass transport and dissipation is given as
\begin{equation}
    v_{\rm acc} = \left( \frac{\xi_a \,  G \, \dot{M}_{\rm g,acc} \,Q^{\rm 1+n} \, \gamma_{\rm diss}}{6(1+\beta)\,f_{\rm g,Q}^{\rm 1+n}} \right)^{1/3} \; .
    \label{eq:acc_vel}
\end{equation}
Here, $\xi_{\rm a}$ is the accretion-driven turbulence efficiency which we assume to be 0.6, $\gamma_{\rm diss}$ is the parameter {representing} turbulence dissipation (assumed as 1 for high redshifts), $\beta$ is a parameter related to the shape of the rotation curve ( 0 for flat rotation curve), and {$Q$} is the {Toomre} parameter which quantifies the stability of rotating, self-gravitating systems like a galactic disc
\begin{equation}
    Q = f_{\rm g,Q} \sqrt{2(1+\beta)} \delta^{-1} \frac{\sigma_{\rm g}}{V_{\rm d}} \; .
\end{equation}
Here, $\sigma_{\rm g}$ is the dispersion velocity related to the circular velocity of disc, $V_{\rm d}$ is the rotation velocity, $\delta$ is the ratio of gas mass within the disc to the total mass within a sphere of radius ($R_{\rm d}$) of the bulge-less disc \citep{2022Ginzburg,2009dekel}, and $f_{\rm g,Q}$ is the effective gas fraction for Toomre's $Q$ which is {approximately} 0.7 for high redshift galaxies having highly enriched gas. We assume constant values based on \cite{2022Ginzburg}. \\

\subsubsection{Turbulent motions in galactic discs}
We consider the net contribution of these two major turbulent velocity contributors, {in quadrature:}
\begin{equation}
    v_{\rm net} = \sqrt{c_{\rm s}^2 + v_{\rm acc}^2 + v_{\rm SN}^2} \; .
    \label{eq:v_net}
\end{equation}
In order to see the effects of turbulent velocity in the model, we replace $c_{\rm s}$ with $v_{\rm net}$ in evaluation of scale height and gas density.

Figure \ref{fig:vturb} shows the variation of turbulent velocities with respect to halo mass for different redshift ranges. The turbulent velocity from accretion increases as the halo mass increases. The turbulent velocity from the supernova feedback initially increases with the halo mass but starts to decline after reaching a maximum at a halo mass of $\sim 10^{10} \rm M_\odot$. This is because low-mass galaxies have a higher fraction of SNe energy relative to binding energy. The solid lines in the figure represent the net turbulent velocity. In this study, we focus on a homogeneous medium, deferring the exploration of non-homogenous medium aspect to future investigations.

\begin{figure}
        \centering
        \includegraphics[scale=0.6]{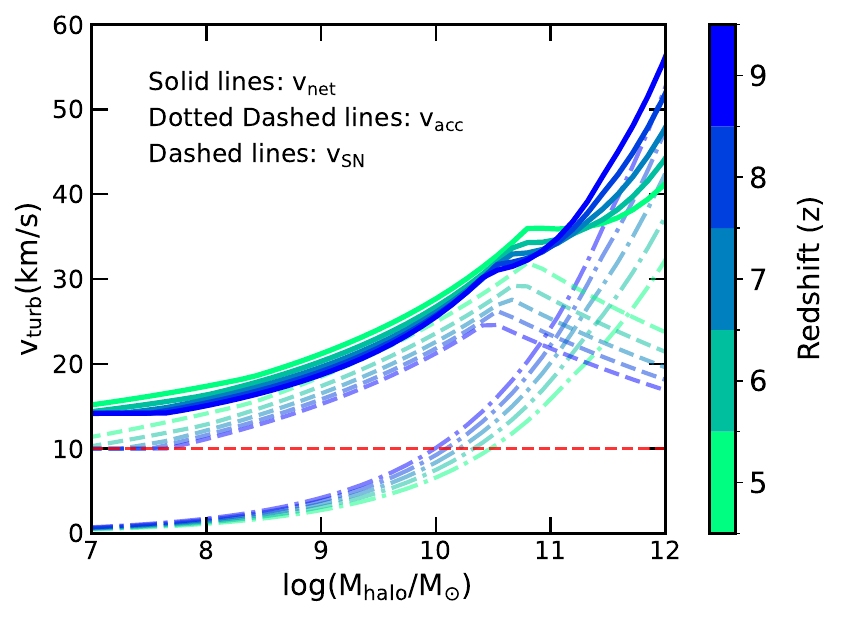}
        \caption{{\rss }The variation of different turbulent velocities with the halo mass (M) for five different redshifts shown in color bar. The dashed lines correspond to the turbulence velocity because of SN feedback $(v_{\rm SN})$ and the dotted-dashed lines refers the turbulence velocity supported by accretion $v_{\rm acc}$. Solid lines represent the net turbulence velocity $(v_{\rm net})$ calculated as a combined effect of both $(v_{\rm SN})$ and $v_{\rm acc}$. The dashed line shows ambient sound speed at $10^4$ K.}
        \label{fig:vturb}
    \end{figure}

\subsection{Metallicity}
\label{sec:metal}
Because our analytic model does not follow an individual galaxy across time to estimate galactic chemical enrichment we adopted a closed box model. In this { limit}, a galaxy does not have inflow or outflow of gas beyond the halo virial radius \citep{1958metallicity,1959schmidt}. Consequently, gas initially present within the galaxy is assumed to remain inside the galaxy with a fraction subsequently forming stars. The metals generated in this process are presumed to be reintroduced into the same interstellar medium (ISM) through enrichment. The enriched gas is then recycled to form new stars.

The metallicity { $Z(t)$} at any stage of galaxy evolution in this model is given as
{
\begin{equation}
    Z(t) = -p \, \ln \left[\frac{M_{\rm g}(t)}{M_{\rm g}(0)}\right],
    \label{eq:metal}
\end{equation}
}
where $p$ is the population yield, $M_{\rm g}(t)$ is the amount of gas present at time $t$ inside the galaxy and $M_{\rm g}(0)$ is the initial mass of gas present. In our analysis, we have implemented equation (\ref{eq:metal}) as
{ 
\begin{equation}
     Z = -p \, \ln\left[\frac{m_{\rm d} M_{\rm halo} - M_*}{m_{\rm d} M}\right].
     \label{eq:Z}
 \end{equation}
 }
Here, $m_{\rm d} M_{\rm halo}$ is the initial disc mass of the galaxy, and ($m_{\rm d} M_{\rm halo} - M_*$) is the residual gas mass in galaxy after the necessary star formation episode, and where $M_*$ is the total stellar mass within the galactic disc:
{ 
 \begin{equation}
     M_* = m_{\rm d} \, f_{*,{\rm tot}} \, M_{\rm halo} \; .
\end{equation}
}
Here $f_{\rm *,{tot}}$ is the total star formation efficiency calculated as { 
 \begin{equation}
f_{*,{\rm tot}} = N_{\rm mergers} \, f_* \; .
\end{equation}
}\\

The population yield, $p$ in equation (\ref{eq:Z}) is estimated by comparing with the best-fit estimates at z = 7 of mass-metallicity relation from \cite{2024Chemerynska} 
\begin{equation}
    12 + \log(O/H) = 0.39^{+0.02}_{-0.02} \times 
    \log(M_*) + 4.52^{+0.17}_{-0.17}.
\end{equation}
Figure \ref{fig:best_fit} shows the variation of 12 + $\log(O/H)$ against the stellar mass $\log_{10}(M_*)$ for different redshifts. 

The blue data-points represent samples of high-redshift JWST galaxies from \cite{2024Chemerynska} while red squares and green diamonds are the average and median values of JWST samples mentioned in  \cite{Nakajima2023} and \citep{2024curti}. For the conversion of total metallicity to an observable gas-phase metallicity, i.e ($12 + \log(O/H)$), we assume oxygen to be {$43 \%$ of the total metal mass and hydrogen approximately $71.5 \%$ of the total gas mass based on \cite{2009Asplund} estimates. }
 
With these assumptions, the population yield used in our analysis is $p \sim 0.016$. Because stellar mass builds up linearly in time during the star-formation episode we expect there to be scatter in the metallicity at fixed SFR. We derive this value based on our metallicity formula (Equation \ref{eq:metal}), considering a total star formation efficiency ($f_{\rm *,tot}$) of 0.2 and applying a $\pm 1 \sigma$ scatter in metallicity across the full range of stellar mass ($M_{\rm *}$), varying from $0\%$ to $100\%$. 
This approach allows us to determine the scatter in metallicity at a fixed SFR.

This variation in stellar mass at fixed SFR introduces scatter of approximately 0.45 dex in $12 + \log(O/H)$ at fixed SFR. We appreciate the fact that the closed-box model may not represent a realistic enrichment scenario, and in particular does not account for any gas exchange including outflows. However, we note that our calibrated relation agrees well with the mass-metallicity measured across the full mass range in \citep{Nakajima2023,2024curti}. In later section, we explore the effect of this scatter on the LF and flux.

\begin{figure}
    \centering
    \includegraphics[scale=0.6]{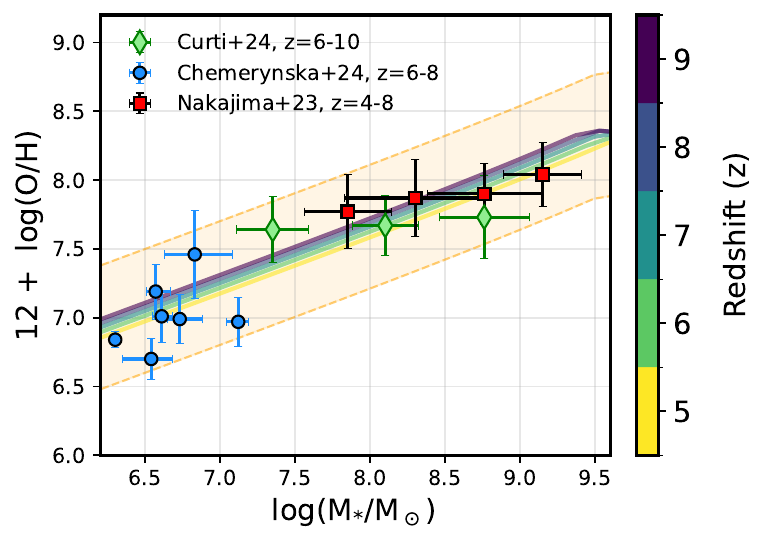}\par
    \caption{Metallicity is plotted against the stellar mass for different redshifts represented by different colored lines. The blue circles represent the samples of high-redshift galaxy candidates from \protect\cite{2024Chemerynska}, the red squares correspond to the avereage value of JWST samples from \protect\cite{Nakajima2023} and the green diamonds correspond to the median values with standard deviation of JWST samples from \protect\cite{2024curti}. The orange dashed curve show the estimate of scatter 0.45 dex in the relation.
    }
    \label{fig:best_fit}
\end{figure} 

\begin{figure}
    \centering
    \includegraphics[scale=0.6]{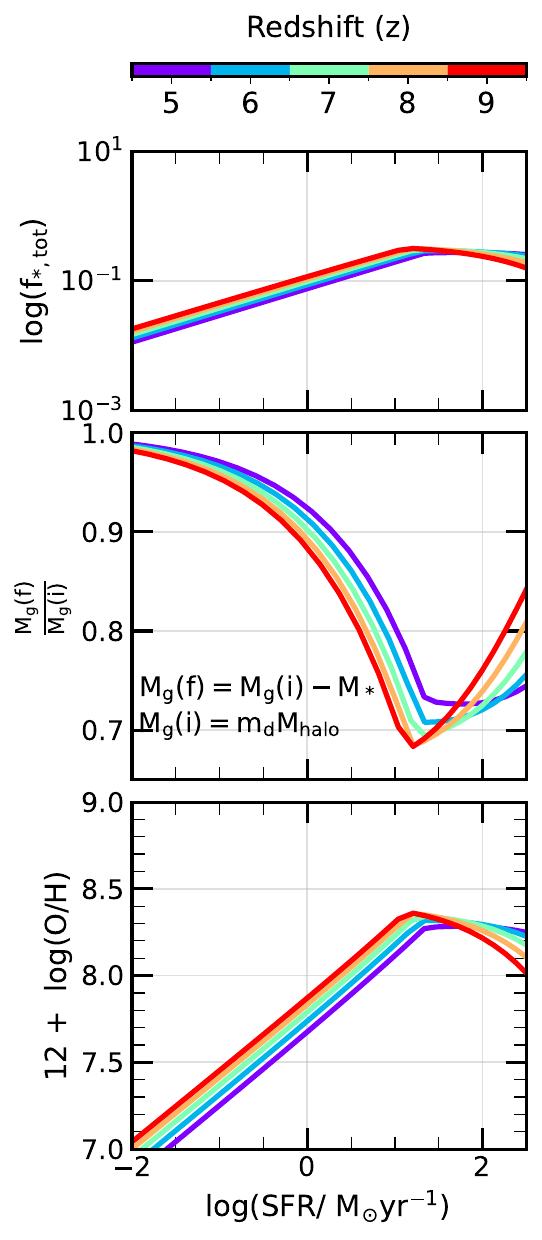} \par
    \caption{Top panel: Star Formation Efficiency ($f_{*,{\rm tot}}$) plotted against Star Formation Rate $\rm log(SFR \ / \ M_{\odot}yr^{-1})$ for five different redshifts. Middle panel: variation of the gas ratio within the disc before and after star formation with respect to SFR.  Bottom panel: The gas-phase metallicity is shown with respect to the $\rm log(SFR \ / \ M_{\odot}yr^{-1})$ for five different redshifts.}
    \label{fig:abundance}
\end{figure}

The calculated metallicity depends upon the amount of gas we have before and after the star formation episode. Therefore, as the total star formation efficiency, $f_{*,{\rm tot}}$ increases, the availability of gas for star formation within the galactic disc decreases. The middle panel of Figure \ref{fig:abundance} shows how the ratio of gas before and after star formation varies with the SFR. For small halo masses, we have a very small $f_{*,{\rm tot}}$ value and hence the gas ratio is close to unity. But as $f_{*,{\rm tot}}$ increases with SFR, the available gas decreases leading to a drop in gas ratio. At large SFR, the ratio increases again after reaching a minimum value. This is because although $f_{*,{\rm tot}}$ remains almost constant (top panel of Figure \ref{fig:abundance}) the larger gas reservoir increases the gas ratio. It is also evident from Figure \ref{fig:abundance} that high redshift galaxies are predicted to have higher value of $f_{*,{\rm tot}}$, which is because of the large number of mergers at high redshift that trigger star formation in our model.\\
The bottom panel of Figure \ref{fig:abundance} shows oxygen abundance $12 + \log(O/H)$ variation with SFR for five redshifts ($z \sim 5,\,6,\,7,\,8,\,\mbox{and}\, 9$). The gas--phase metallicity initially increases linearly with the star formation rate (SFR) for all redshifts, attains a maximum near $\rm log(SFR)  \sim 1  \ M_{\odot}yr^{-1}$ and is approximately constant for larger SFR. High-z galaxies in our model have large abundances at fixed SFR compared to smaller redshifts for all but the largest SFR. This counter-intuitive result arises from the scaling of metallicity in our closed-box model (Equation \ref{eq:Z}). From the SFR-halo mass relation, it is evident that for a fixed SFR, galaxies reside in lower-mass halos at higher-z. Since the metallicity in Equation \ref{eq:Z} depends on the total stellar mass and the initial gas mass (which scales with $M_{\rm halo}$), for a fixed SFR, galaxies at higher redshifts have lower $M_{\rm halo}$, meaning that the gas reservoir $m_{\rm d}M_{\rm halo}$ is smaller. This results in a larger fraction of metals being retained in the ISM, leading to higher metallicity in our model.

\subsection{Ionization Parameter}
The ionization parameter $q$ (measured in cm/s) is defined as the ratio of the mean ionizing photon flux $(\phi_{\rm HI})$ to the number density of the hydrogen atoms $({n_{\rm p}})$ \citep{2003Doita,2019Lisa}
{ 
\begin{equation}
    q = \frac{\phi_{\rm HI}}{n_{\rm p}},
    \label{eq:IP1}
\end{equation}
}
where $n_{\rm p}$ is calculated from equation (\ref{eq:rscale}).

{Several} methods have been introduced to calculate the ionization parameter, including at the inner edge of a plane parallel nebula \citep{Kewley2002} and the volume averaged ionization parameter \citep{Stasinska2015}. We use volume--weighted mean ionization parameter from \citep{2015Kewley} 
\begin{equation}
    q = \frac{2^{2/3} Q_*}{4 \pi R_{\rm s}^2 n_{\rm p}} \; ,
    \label{eq:Ionpar1}
\end{equation}
where $R_{\rm s}$ is the Str\"omgren radius
\begin{equation}
    R_{\rm s} = \left(\frac{3Q_*}{4 \pi \alpha_B n_{\rm p}^2}\right)^{1/3}
    \label{eq:Rs}
\end{equation}
Here, $\alpha_B$ is the case B recombination coefficient having a value $\sim 2.6 \times 10^{-13} \mbox{cm}^3 \mbox{s}^{-1}${, at 10$^4$\;K \citep{1995Hummer}.}

The symbol $Q_*$ in Equations (\ref{eq:Ionpar1}) and (\ref{eq:Rs}), represents the ionizing photon luminosity (photons/sec). 
{
We calculate $Q_*$ using the following steps:}
\begin{itemize}
    \item We assume that 10 supernovae collectively form a superbubble of radius $R_{\rm e}$ (eq. \ref{eq:R_e}). Our results are not qualitatively dependent on this number.
    
    \item We then utilize the superbubble volume and the mass density of particles {$(n_{\rm p} \,\mu\, m_{\rm p})$} to compute the mass of gas enclosed within the superbubble ($M_{\rm bub}$). This allows us to calculate the number of superbubbles within the galactic disk, denoted as $N_{\rm bub}$, using:

    { \begin{equation}
        N_{\rm bub} = \frac{\epsilon \ (m_{\rm d}  M)}{M_{\rm bub}} \; .
    \end{equation}
    }
    Here, $m_{\rm d} M_{\rm halo}$ is the disc mass and $\epsilon$ is the porosity which we {take} as
    \begin{equation}
       \epsilon = \frac{SFR_{\rm crit}}{SFR} \;
       \label{eq:porosity}
    \end{equation}
    We incorporate a condition such that if $SFR$ exceeds $SFR_{\rm crit}$, then $\epsilon$ equals 1.
    \item The next step involves using $N_{\rm bub}$ to calculate ionizing flux per bubble. 
    
    We assume a value of $N_\gamma \approx 4000$, ionizing photons produced per baryon \citep{Barkana2001}. We use this quantity to determine the ionizing photons per second, denoted as $Q_* = \frac{N_\gamma \ SFR}{N_{\rm bub} \ m_{\rm p}}$, { used in} subsequent ionization parameter calculations.
    
\end{itemize}

\section{Effect of turbulence on galaxy physical properties}
\label{sec:results}

Figure \ref{fig:SFRD_comp} shows the effect of turbulence on the Star Formation Rate Density (SFRD) function. We consider the same value for $f_{*,max}$ = 0.035 for turbulent and non-turbulent cases but different starburst lifetimes, $t_{SF}$ = $2.0 \times 10^7$ years \ / $2.5 \times 10^7$ years for turbulent / non-turbulent cases. The figure shows that there are only minimal differences observed at higher SFR values. 

\begin{figure}
    \centering
    \includegraphics[scale=0.6]{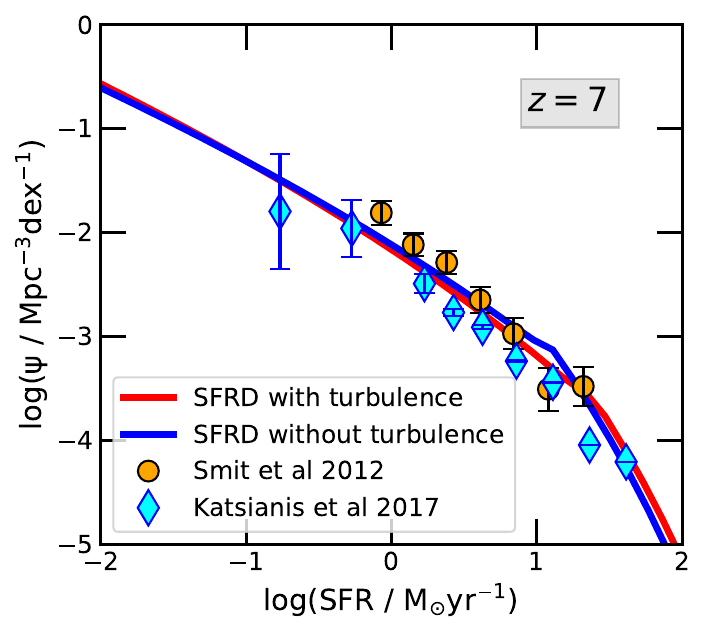}\par
    \caption{{\rss }SFRD function for turbulent (shown by red line) and non-turbulent (shown by blue line) cases for redshift $z \sim 7$. The cyan colored points correspond to the results of \citep{2017Katsianis} and orange points represents the observations of \citep{Smit2012}.}
    \label{fig:SFRD_comp}
\end{figure}

\begin{figure*}
    \centering
    \includegraphics[scale=0.57]{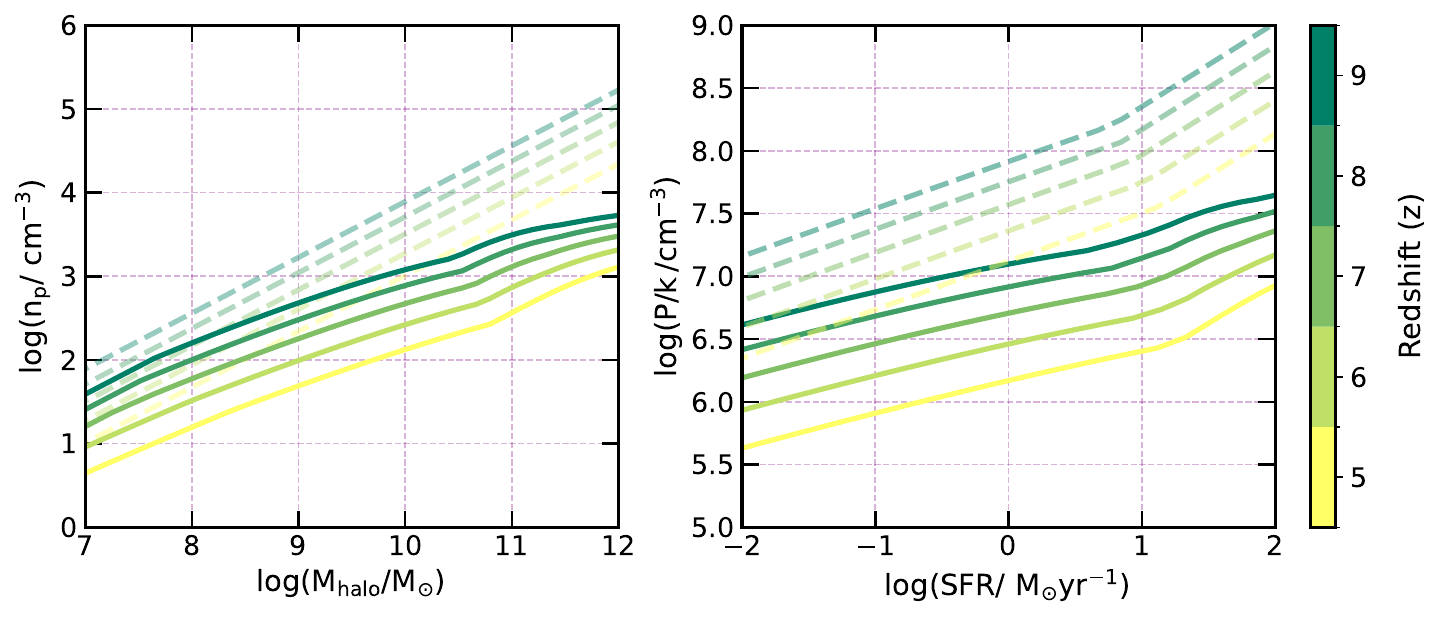}\par
    \caption{Left panel: Number density ($\rm cm^{-3}$) is plotted against the halo mass $M_{\rm halo}$. Right panel: The pressure $\log_{10}(P/k)$ plotted against the SFR inside a $\HII$ region. Curves are shown for five different redshifts $z \sim 5,6,7,8, 9$. Solid and dashed lines in both panels correspond to the turbulent and non-turbulent discs respectively.}
    \label{fig:npd}
\end{figure*}

\begin{figure*}
        \centering
        \includegraphics[scale=0.59]{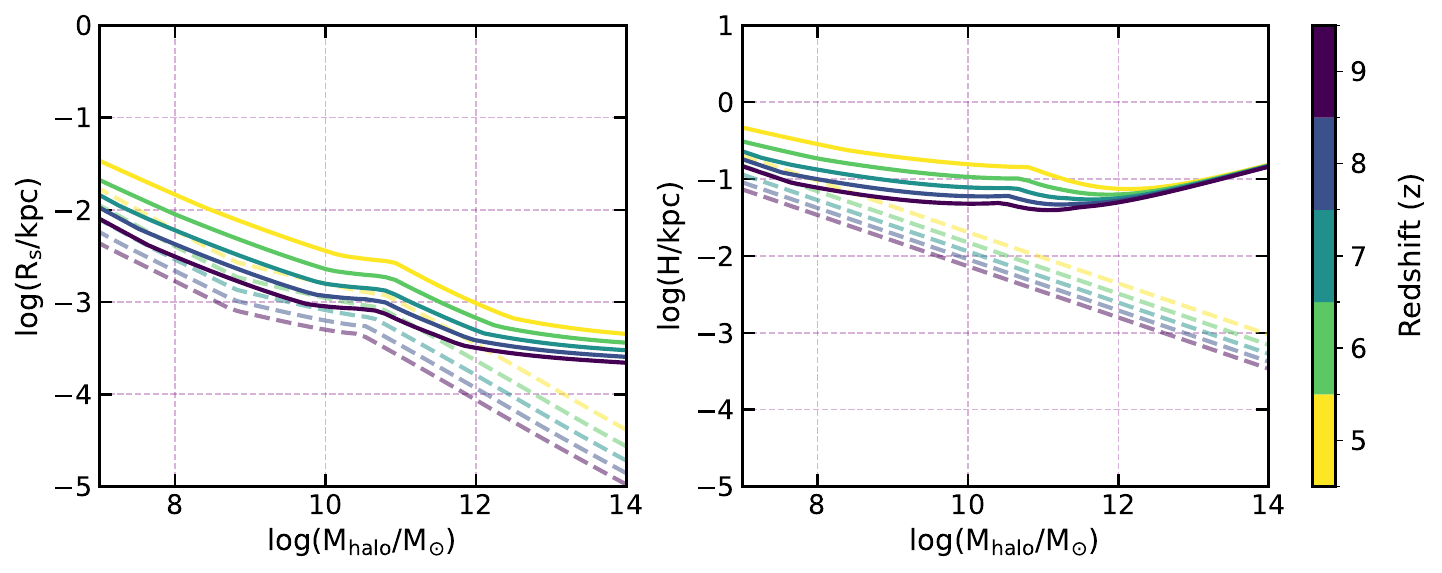}
        \caption{{ The variation of Str\"omgren} radius ($R_{\rm s}$) and scale height of disc (\rm H) with respect to halo mass for five different redshifts in the left and right panels respectively. The solid lines correspond to the case when turbulence is taken into account and dashed lines are for the non-turbulent case. }
        \label{fig:scale_height}
\end{figure*}
The density of gas is a critical parameter for ionization rate. Figure \ref{fig:npd} presents the variation of $n_{\rm p}$ with respect to halo mass for different redshift ranges. The solid lines in the figure correspond to the number density in a turbulent galactic disc, while the fainter dashed lines represent values for a non-turbulent galactic disc, as originally assumed in the model. Equation (\ref{eq:rscale}) results in increased gas density with respect to halo mass. Furthermore, high-redshift galaxies tend to exhibit higher gas densities as a consequence of their increased compactness at earlier cosmic epochs. However, the presence of turbulent support within the galactic disc significantly reduces the value of $n_{\rm p}$ relative to a thermally supported disc.

Figure \ref{fig:npd} also shows pressure, $P$, as a function of SFR assuming a constant local gas temperature within the $\rm HII$ region of $10^4$\,K  \citep{2019Lisa}, following the equation:
\begin{equation}
    P = n_{\rm p} \, k\, T,
\end{equation}
where $k$ is the Boltzmann constant. We note that this explicitly assumes turbulent motion decreases average density due to bulk motions but does not increase local thermal gas temperature. The right-hand side of Figure \ref{fig:npd} demonstrates that a higher star formation rate, which occurs in larger halos is associated with increased pressure within the galactic disc. Additionally, pressure in galactic discs increases at high redshifts. Since pressure is proportional to the number density, { there is a reduction in gas thermal pressure due to turbulence within the galactic disc at a given SFR.\\

The left and right panels of Figure \ref{fig:scale_height} depict the variation of the Str\"omgren sphere radius ($R_{\rm s}$) and the scale height of the disc ($H$) with respect to halo mass ($M$) for different redshifts respectively. The light dashed colored lines correspond to the scale height of the disc in the absence of turbulence, while the solid lines represent the disc's scale height with the inclusion of turbulence.
Both the Str\"omgren radius and the scale height of a turbulent disc are significantly larger than for a non--turbulent disc, especially for larger halo masses. Notably, in the large halo mass range, the scale height's dependence on redshift is predicted to be minimal. This arises due to the redshift dependencies of different parameters used in the scale height calculation. The scale height ($H$) has a $(1+z)^{-2}$ dependence, as outlined in equation (\ref{eq:H}). Additionally, the net turbulent velocity ($v_{\rm net}$), as specified in Equation \ref{eq:v_net} which is dominated by accretion for large mass follows $v_{\rm acc} \propto (1+z)^{5/6}$. Consequently, {in} the high halo mass range, the overall redshift dependence of $H$ varies as $(1+z)^{-1/3}$, resulting in {only slow} variations of $H$ with redshift. Later in the paper, we describe how the dependence of $R_{\rm s}$ on turbulence can influence properties including ionization parameter.

\subsection{Ionization Parameter }

In Figure \ref{fig:IP}, we show ionization parameter variation with SFR for five different redshifts. The dashed lines correspond to the ionization parameter without turbulence and solid lines correspond to the ionization parameter with turbulence. In the local universe, the ionization parameter mostly lies within the range of {$7.28<\log(q)<7.58$ } \citep{2019Lisa} which is significantly lower than predicted at higher redshift (see also \citealp{2018Lisa}). We predict a very high ionization parameter (greater than $10^9 \ \rm cm/s$)  in our non-turbulent case because of the high gas density these galaxies posses at high redshift in our model. We see this behavior from inspection of equations (\ref{eq:npd}-\ref{eq:R_vir}) and (\ref{eq:Ionpar1}-\ref{eq:Rs}) which imply 
\begin{equation}
    q \propto n_{\rm p}^{1/3} \propto (1+z)^{4/3}.
\end{equation}
For the turbulent case, the ionization parameter is also higher than at low redshift z. However, we note that in both the turbulent and non-turbulent cases extrapolation of ionization parameter in Fig \ref{fig:IP} to $z \sim 0$ yields values of log(q)$\sim 7-7.5$ in agreement with low-z measurements. Thus, the predicted redshift evolution of ionization parameter is consistent with z= 0  values.

\begin{figure}
    \centering
    \includegraphics[scale=0.58]{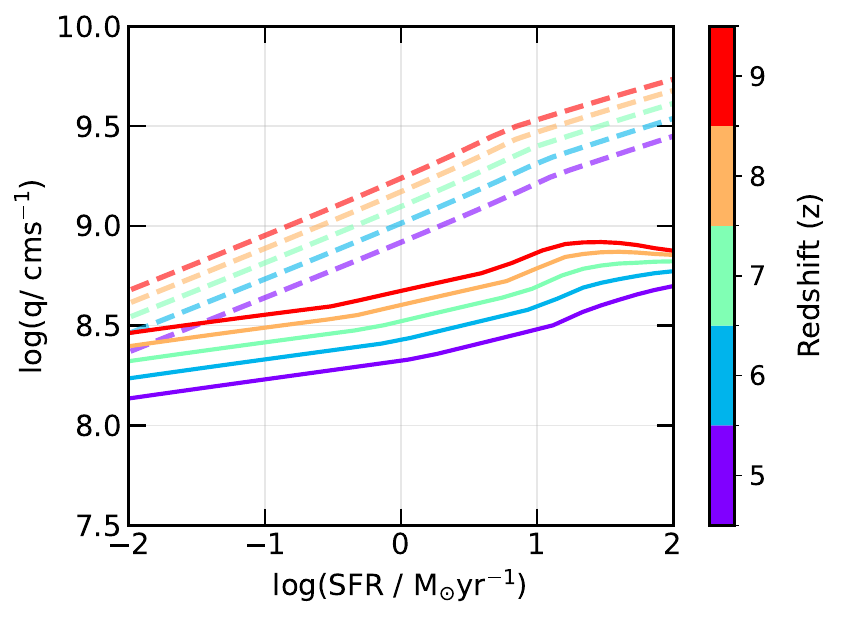}\par
    \caption{{\rss }The variation of ionization parameter with respect to star formation rate for five different redshifts. The solid lines correspond to ionization parameter in a turbulent galactic disc while dashed lines refer to a non-turbulent disc. }
    \label{fig:IP}    
\end{figure}
\FloatBarrier

\section{Emission line luminosity}
\label{sec:emission_line}

\begin{figure}
    \centering
    \includegraphics[scale=0.6]{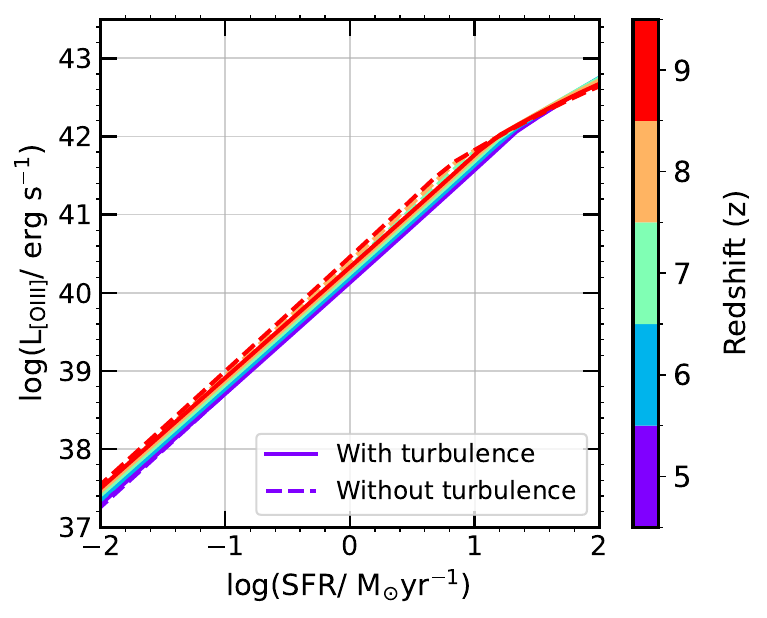}
    \caption{The variation of line luminosity with respect to star formation rate, $\rm log(SFR \ / \ M_{\odot}yr^{-1})$ for five different redshifts: $z \sim 5,\,6,\,7,\,8 \ \mbox{and} \ 9$, shown in the color bar. The solid and dashed lines correspond to turbulent and non-turbulent cases.}
    \label{fig:[OIII]_luminosity}
\end{figure}

The {\OIII\ } emission line is known to depend on physical properties including ionization parameter and metallicity \citep{2008Brinchmann,2013Kewley,Kewley_2015,2013aKewley,Masters_2014}. In this section, we adopt the model of \cite{Lidz20} to compute the { \OIII\ } emission line luminosity within our analytic model.
The {\OIII\ } rest-frame optical emission line is a collisionally excited line. We calculate the line luminosity from the {$^1$D$_2 \rightarrow ^3$P$_2$ } transition corresponding to { $\lambda \sim 5007.8 \ $\AA } \ of {the } doubly ionized oxygen atom. Assuming the oxygen to be mostly doubly ionized, the luminosity of the{ $\OIII \lambda 5007$} emission line is :
\begin{equation}
  \begin{split}
    L_{\OIII} = \left(\frac{n_{\rm O}}{n_{\rm H}}\right)_\odot \  \frac{Z}{Z_\odot}  \ \left(k_{03}+k_{04}\frac{A_{43}}{A_{43}+A_{41}}\right)\frac{A_{32}}{A_{32}+A_{31}} \\
    \times \ h\nu_{\rm 32} \left(\frac{Q_{\HI}}{\alpha_{\rm \beta,\HII}}\right) \  \left(\frac{V_{\rm \OIII}}{V_{\rm \HII}}\right)\; .
  \end{split}
  \label{eq:L32}
\end{equation}  
Here, $n_{\rm O}/n_{\rm H}$ is the solar oxygen abundance with respect to hydrogen, $\frac{Z}{Z_\odot}$ is the metallicity in solar units, $h\nu_{32}$ is the energy associated with the $\OIII,  \lambda 5007$ emission line and $Q_{\HI}$ is the ionizing photon luminosity for the disc (photons/s). The $A_{\rm ul}$ are the Einstein spontaneous coefficients whose values have been adopted from \citep{1996Wiese}. The $k_{\rm lu}$ are the collisional excitation coefficients given as 
\begin{equation}
    k_{\rm lu} = \frac{\beta}{\sqrt{T}} \  \frac{\Omega_{\rm lu}}{g_l} \ e^{-\left(E_{\rm lu}/k T\right)},
    \label{eq:klu}
\end{equation}
 where $E_{\rm lu}$ is the excitation energy, $g_l$ is a statistical weight, and $\Omega_{\rm ul}$ is collisional strength values mentioned in \citep{2011Drain}. The $\beta$ used in equation (\ref{eq:klu}) is 
 \begin{equation}
     \beta = \left(\frac{2 \pi \hbar^4}{k m^3}\right)^{1/2},
 \end{equation} 
where {$m$} is mass of electron and $k$ is the Boltzman constant. Table \ref{tab:coeff} shows the various coefficient values used in equation (\ref{eq:L32}). 

The term $V_{\rm \OIII}/ V_{\rm \HII}$ represents the fraction of doubly-ionized oxygen averaged over the $\HII$ region. This factor becomes lower than unity at low values of both $n_{\rm p}$ and $Q_{\rm HI}$. However, in the regime predicted for high-z galaxies with log$(n_{\rm p}) \gtrsim 2$ and log$(Q_{\rm HI}) \gtrsim 50$, the correction term is approximately $Q \geq 0.9-0.95$ see \citep{Lidz20}, which has a negligible effect on our calculation of line luminosity. We therefore consider the volume correction term to be 1.\\

\begin{table} 
\begin{tabular}{ p{1.0cm} p{4cm} p{2.5cm}}
\hline
\\
$u \rightarrow l$ & $\Omega_{\rm ul}$ & $A_{\rm ul} \; \mbox{(cm$^3$/s)}$ \\
\hline
\hline
\\
$3 \rightarrow 0$ & $0.243 (2J+1) \times T_4^{0.120 \ + \ 0.031 \,\ln T_4}$ & $A_{31} = 4.57\times 10^{-6}$ \\

                 & (fit to \cite{1999Aggarwal}) & $A_{32}= 3.52\times 10^{-5}$ \\
\hline
\\
$4 \rightarrow 0$ &  $0.0321 (2J+1) \times T_4^{0.118 \, + \, 0.057 \, \ln T_4}$ & $A_{41}=2.5\times 10^{-1}$ \\ 
       &   (fit to \cite{1999Aggarwal}) & $A_{41}=1.7\times 10^0$  \\
\hline
\end{tabular}
\caption{Coefficients used in the derivation of $\OIII \lambda 5007$ line. First column: Transitions from upper level ($u$) to lower level ($l$). Middle column : Collisional strength of the line as mentioned in \citep{2011Drain}. In our case, we consider $J=0$. Third column: Einstein spontaneous coefficient (in $\rm cm^3/s$)}
\label{tab:coeff}
\end{table}

 Figure \ref{fig:[OIII]_luminosity} illustrates the predicted [OIII] $\lambda 5007$ emission line luminosity as a function of Star Formation Rate (SFR) across various redshifts. Our model predicts a small redshift-dependent evolution in line luminosity. This behaviour stems from the inherent one-to-one relationship between line luminosity and SFR within our model. The figure also suggests that the evolution of line-luminosity with SFR has only a small dependence on turbulence.\\

\begin{figure}
    \centering
    \includegraphics[scale=0.6]{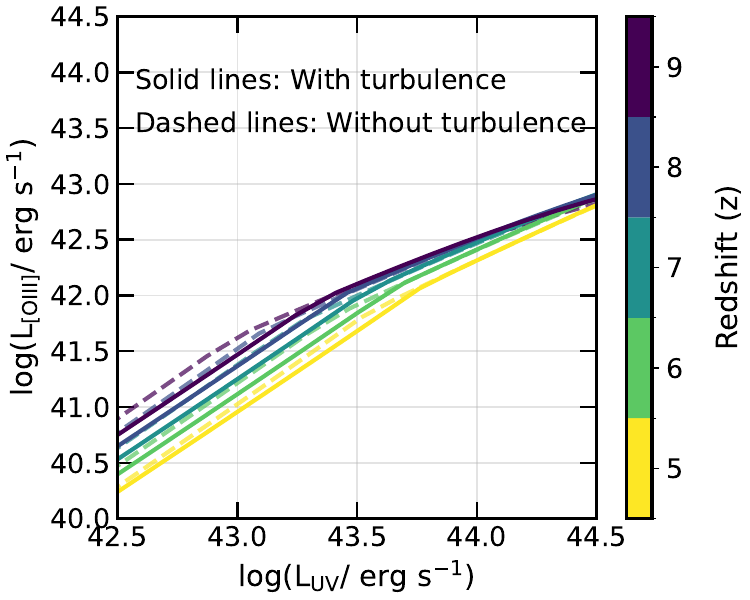}
   \caption{The variation of [OIII]5007 line luminosity with the UV luminosity for five different redshifts. The solid and dashed lines correspond to turbulent and non-turbulent cases respectively.}
    \label{fig:LUV_LOIII}
\end{figure}

\subsection{\texorpdfstring{$\OIII$}{OIII}-UV luminosity relation}

\label{sec:OIII_UV}
We investigate the relationship between the [OIII]$\lambda5007$ line luminosity and the UV line luminosity. The UV luminosity is derived from our modeled Star Formation Rate (SFR) following the relationship of \citep{Madau1998}

\begin{equation}
    L_{\text{UV}} = 8.0 \times 10^{27} \times \frac{\text{SFR}}{\rm M_\odot \text{ \rm yr}^{-1}} \, \text{ \rm ergs} \, \text{s}^{-1} \, \text{\rm Hz}^{-1}\; ,
    \label{eq:LUV}
\end{equation}
where the constant is assumed for a Salpeter Initial Mass Function (IMF) at a wavelength of 1500 \,\AA \ and the luminosities at this wavelength have been averaged over a rectangular bandpass of width $\Delta \lambda / \lambda = 20 \%$ \citep{Madau1998}. Figure \ref{fig:LUV_LOIII} illustrates the relationship between the line luminosity $L_{\OIII}$ and the UV luminosity $L_{\rm UV}$. The solid and dashed lines in the figure represent turbulent and non-turbulent scenarios. 

We compare the modeled variation in the \OIII/UV luminosity ratio with $L_{\rm UV}$ from our analysis to measurements of this relation from FRESCO \citep{2024Meyer,2023Pascal} for redshifts around $z \sim 7$ and $z \sim 8$ in Figure \ref{fig:comp_LUV_LOIII}. The scatter in \OIII/UV luminosity ratio can be attributed to variations in the gas content within the disc. The scatter of 0.45 dex in $12 + \log(O/H)$ discussed in section \ref{sec:metal} corresponds to scatter in $\OIII$ line luminosity that is also 0.45 dex. Figure \ref{fig:comp_LUV_LOIII} shows that the values, trend and scatter predicted by the modelling (which is calibrated against the SFRDF at $z \sim 6$) agree well with these first systematic observations of high-z emission line galaxies. 

\begin{figure}
    \centering
    \includegraphics[scale=0.65]{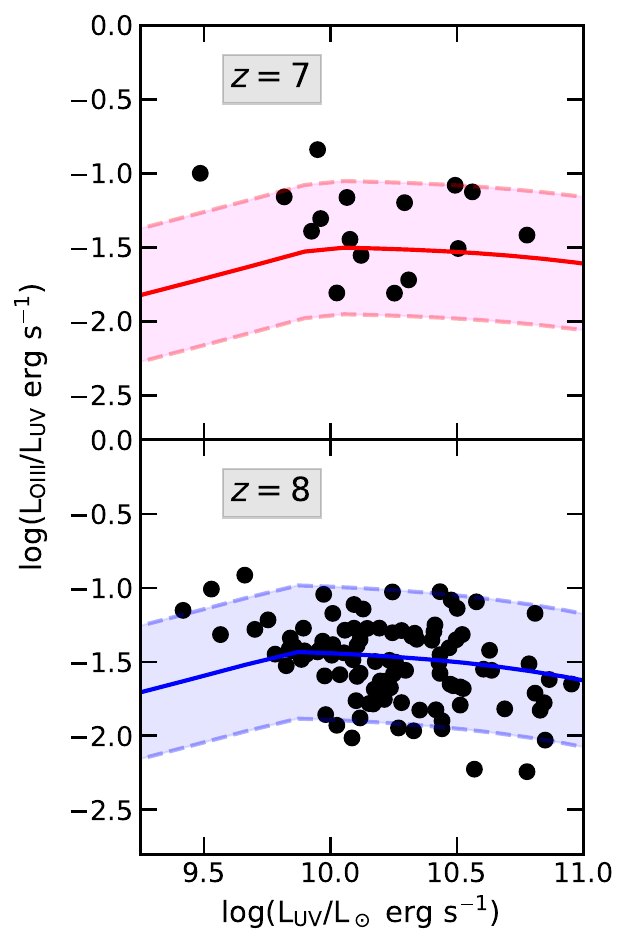}
   \caption{The predicted \OIII /UV luminosity ratio as a function of UV luminosity for $z \sim 7$ and 8 (solid lines). The FRESCO measurements are shown with black circles in both panels \citep{2024Meyer}. The shaded area in both panels corresponds to 0.45~dex scatter in \OIII\ line luminosity.}
    \label{fig:comp_LUV_LOIII}
\end{figure} 

\FloatBarrier
\section{Luminosity Function and number counts}
\label{sec:LF}

In this section, we derive the [OIII] luminosity function from the SFRD function $\phi(SFR)$ (equation \ref{eq:SFRD}), the mean relation between SFR and $\OIII$ luminosity $\bar L_{\OIII}$ and the scatter in the SFR-$L_{\OIII}$ relations. We have

\begin{equation}
\begin{split}
    \phi(\log \bar L_{\OIII}) = \int_{-\infty}^{\infty} \frac{dn}{d\log \bar L_{\OIII}} \  N(\log L_{\OIII}|\log \bar L_{\OIII}) \\
     \times \ d\log L_{\OIII},
\end{split}    
\end{equation}\\
where, $N(L_{\OIII}|\log \bar L_{\rm [OIII]})$ is given as 

\begin{flalign*}
N(L_{\OIII} | \log {\bar L_{\OIII}}) & = 
\frac{1}{\sqrt{2\pi\sigma^2}} \\
& \times \exp\left( -\frac{(\log L_{\OIII} - \log { \bar L_{\OIII}})^2}{2\sigma^2} \right),
\end{flalign*}
and $\sigma =$ 0.45 dex scatter in the SFR-$L_{\OIII}$ relation. The term $dn/d\log \bar L_{\OIII}$ corresponds to the luminosity function with no scatter evaluated as 

\begin{equation}
    \frac{dn}{d\log( \bar L_{\OIII})} = \phi(SFR)\,\frac{d\log(SFR)}{d\log( \bar L_{\OIII})}
    \label{eq:LFequation}
\end{equation}\\
where $\phi(SFR)$ is in $\rm Mpc^{-3}dex^{-1}$ and $L_{\OIII}$ is the luminosity of $\OIII \lambda5007$ emission line from equation (\ref{eq:L32}). 

The resulting LF, $\phi(L_{\OIII})$ is depicted in Figure \ref{fig:LF} (solid lines). Also shown (dashed lines) is the relation without including scatter in the $L_{\OIII}$-SFR relation (equation \ref{eq:LFequation}). Recent observations based on JWST surveys have made it possible to measure the luminosity function of $\OIII \lambda5007$ emission line galaxies. In the figure, orange circles with error bars represent the emission line measurements of galaxies observed in the EIGER (Emission-line galaxies and the Intergalactic Gas in the Epoch of Reionization) survey using the first deep JWST/NIRCam wide field slitless spectrscopy \citep{Mathhee2022}. Also, green squares and blue diamonds in the figure are the LF measured from FRESCO \citep{2024Meyer,2023Pascal}. Comparison of the LF observations with the model predictions illustrate good agreement.

Figure \ref{fig:comp_LF} illustrates a comparison of luminosity functions assuming turbulent and non-turbulent discs at redshift $z \sim 7$ and shows little dependence. Figure \ref{fig:NC_galaxies} shows the number counts of emission line galaxies for the FRESCO survey. We have considered two different redshift ranges $(z \sim 7-8, \ 8-9) $ and the total number of galaxies which can be observed within the redshift range of $(z\sim 7-9)$. The expected observed flux for the $\OIII \lambda 5007$ line for JWST NIRCam/grism observation is $3.3 \times 10^{-18}$\, \rm erg/s/cm$^2$ with a 6-sigma emission line, which is shown by the vertical line. The predicted number counts for this flux is in the range $\sim 315-335$ galaxies. We can compare our model with FLARES \citep{2023FLARES} who find a significantly steeper luminosity function  than our predictions. The FLARES calculation at bright end also deviates from the observational constraints. In contrast, the luminosity function from SPHINX \cite{2023katz} is in good agreement with our model and observations.

\vspace{-\baselineskip}
\section{\texorpdfstring{$\OIII / H_{\rm \beta}$}{OIII / H beta} emission line ratio predictions}

\label{sec:line_ratio}

The doubly ionized oxygen $({\OIII}~\lambda 5007)$ and the hydrogen ($H_{\rm \beta}$) recombination lines play crucial roles as tracers of gas properties in star-forming galaxies. The $({\OIII}~\lambda 5007)$ line is particularly sensitive to the ionization parameter and the gas-phase metallicity, while the $H_{\rm \beta}$ line responds to ionizing radiation. The ratio of these two lines therefore offers valuable insights into radiation and ISM properties, and minimizes potential uncertainties arising from dust absorption due to their small wavelength difference. In this subsection, we investigate predictions for the ratio of these two lines based on our model.

We employ the $\MV$ v5.2.1 photoionization modeling code \citep{1996mappings,2008Allen,2017suttherland} to calculate the flux ratio within a parameter space defined by our model. We consider abundance files corresponding to 3 Myr evolved cluster models.  For the sake of simplicity, we avoid any abundance offset or dust correction while using StarBurst99 abundances. The code uses the \cite{2009Asplund} solar metallicity abundance pattern with Kroupa IMF and photoionization atomic models for calculating the flux ratio. We note here that abundances are not expected to be solar at high-z. However, the Asplund abundance was used to calibrate the mass-metallicity relation in Figure \ref{fig:best_fit}.

We estimate the ${\OIII} / H_{\rm \beta}$ line ratio calculated in the parameter space of physical properties describing high redshift galaxies. Figure \ref{fig:line_diag} shows the ${\OIII} / H_{\rm \beta}$ diagnostic line ratio for four different metallicities in parameter space of ionization parameter and pressure based on the $\MV$ v5.2.1 photoionization modelling code. The solid lines correspond to the modelled value diagnostics from $\MV$ v5.2.1.  

Figure \ref{fig:line_ratio} presents the variation of $\log{\OIII}/H_{\rm \beta}$ flux ratio with the line luminosity $\log(L_{\OIII } \,/{\rm ergs \ s^{-1}})$ 
for different redshifts. As with the deviations observed in physical properties such as pressure, metallicity, and ionization parameter between turbulent and non-turbulent cases, we also note variations in the flux ratio when plotted against line luminosity. The red dots in the figures represents scenario when turbulence is considered. Turbulent discs have lower ${\OIII} /H_{\rm \beta}$ values compared to their non-turbulent counterparts emphasizing the impact of turbulence on the ionization state of the interstellar medium (ISM). The grey horizontal band depicts the measured flux ratios of ${\OIII}$ emitters for the full sample stack as observed in the FRESCO survey \citep{2024Meyer}. Figure \ref{fig:weighted_line_ratio} shows the weighted flux ratio calculated from the model for redshifts $z \sim 7,8,9$ against the line luminosity. The solid green and red colored lines correspond to best fit curves for turbulent and non-turbulent cases. The blue circles with error bars correspond to mean flux ratio of $\OIII$ emitters measured in FRESCO. Figure  \ref{fig:line_ratio} and \ref{fig:weighted_line_ratio} illustrate that the model predicts flux ratios which are consistent with a turbulent ISM while flux ratios from a non-turbulent ISM are predicted to be larger than observed.

The predicted variation in flux ratio for turbulent vs non-turbulent disks may therefore provide an important diagnostic of the turbulent ISM.\\

\section{Model caveats}
\label{sec:cav}
The model described in this paper successfully explains the evolution of various physical properties and makes predictions well within the uncertainity range of the observed luminosity function (LF) and the $\OIII$-UV relation from JWST FRESCO observations at high redshift. However, the model has limitations and is based on several crude assumptions which we summarise before conclusion. Notably, the line luminosity calculation assumes a volume filling fraction of [O III] equal to 1, thereby neglecting contributions from other potential oxygen species. This assumption is justified in light of the \cite{Lidz20} model, where the volume correction factor is close to unity for high-density regions and ionization parameters. Additionally, the gas-phase metallicity is calculated assuming solar abundance and a closed-box model, which will not accurately represent the observed abundances. However the model metallicity is callibrated to reproduce the observed $\rm [O/H]$ mass-metallicity relation observed with JWST.

\begin{figure*}
    \centering
    \includegraphics[scale=0.6]{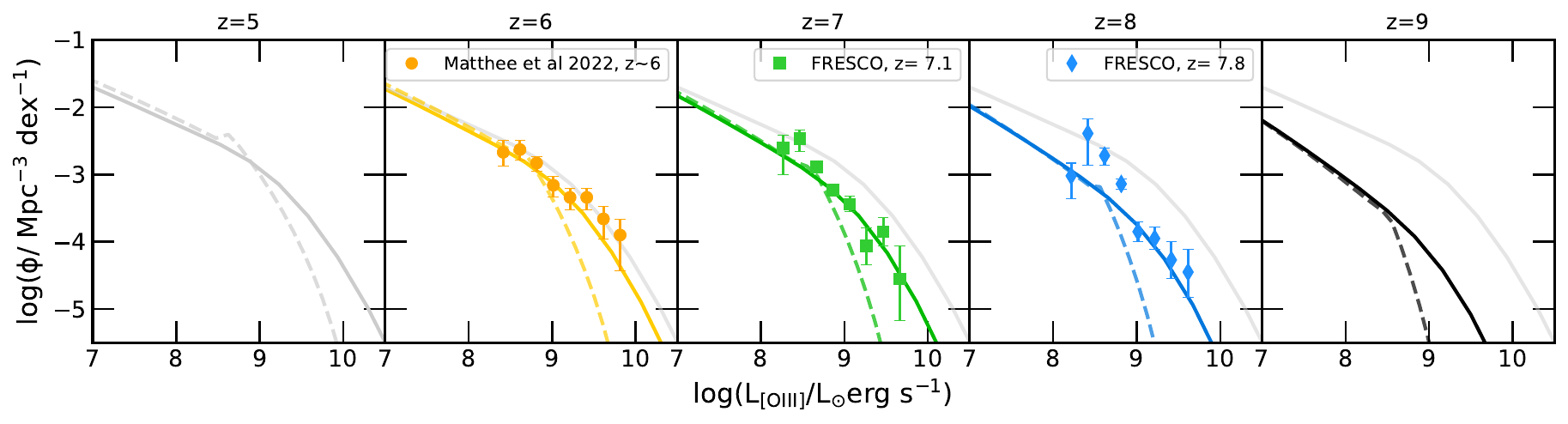}
    \caption{Comparision of LFs for \OIII\ line against line luminosity, $\log(L_{\OIII})$ in units of solar luminosity. Solid lines correspond to scattered LFs including estimate of scatter in $L_{\OIII}$ while the dashed lines refer to LF with no scatter. Orange circles represent the observed LF data for ($\OIII \lambda5007$) emitters corresponding to redshift $z \sim 6$ \citep{Mathhee2022}. Green squares and blue diamonds represent observed {\OIII} emitters from FRESCO survey correspond to redshift $z \sim 7.1$ and $z \sim 7.9$ \citep{2024Meyer}. Light grey lines in last four panels correspond to $z=5$ LF. }
    \label{fig:LF}
\end{figure*}

\begin{figure}
    \centering
    \includegraphics[scale=0.64]{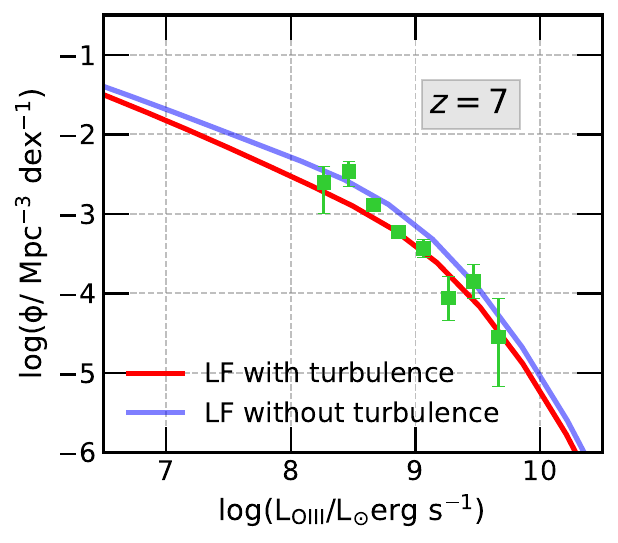}  
    \caption{Comparison of the LF for \OIII\ line plotted against the line luminosity, $\log(L_{\OIII})$ for turbulent and non-turbulent cases. Green squares represent FRESCO $\OIII$ emitters for $z \sim 7$ \citep{2024Meyer}.}
    \label{fig:comp_LF}
\end{figure}

\begin{figure}
    \centering
    \includegraphics[scale=0.5]{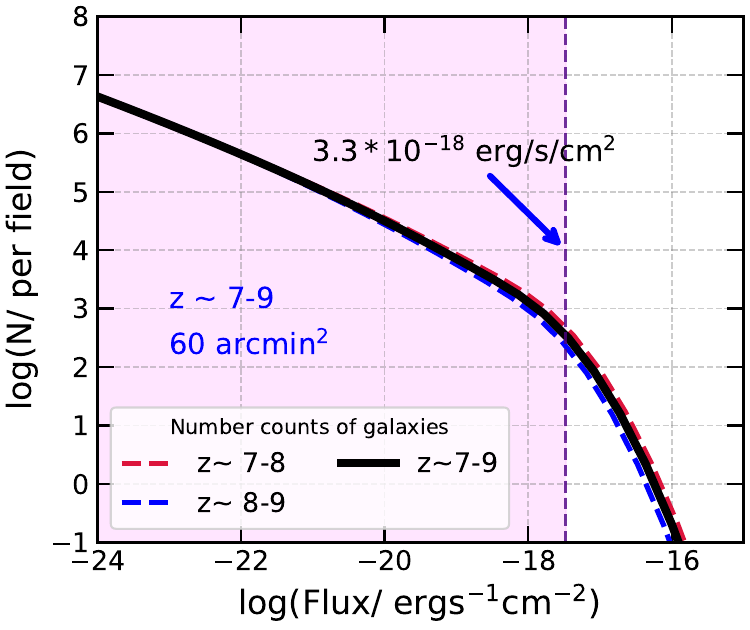}
   
    \caption{The number counts of galaxies for JWST FRESCO survey per field is plotted against the flux (\rm erg/s/cm$^2$). The vertical dashed indigo line shows the $\OIII \lambda5007$ flux limit for NIRCam/grism observations. }
    \label{fig:NC_galaxies}
\end{figure}

\begin{figure}
    \centering
    \includegraphics[scale=0.63]{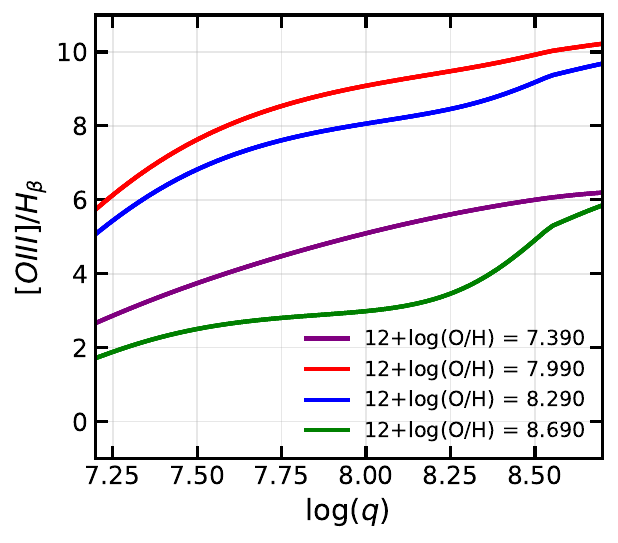}
   
    \caption{$\OIII/H_{\rm \beta}$ is plotted against $\log(q)$ for four different metallicities ($12 + \log(O/H)$= 7.390, 7.990, 8.290 and 8.690 shown by purple, red, blue and green colors respectively) and based on the \MV\ 5.2v photoionization modelling code.}
    \label{fig:line_diag}
\end{figure}

\begin{figure*}
    \centering
    \includegraphics[scale=0.6]{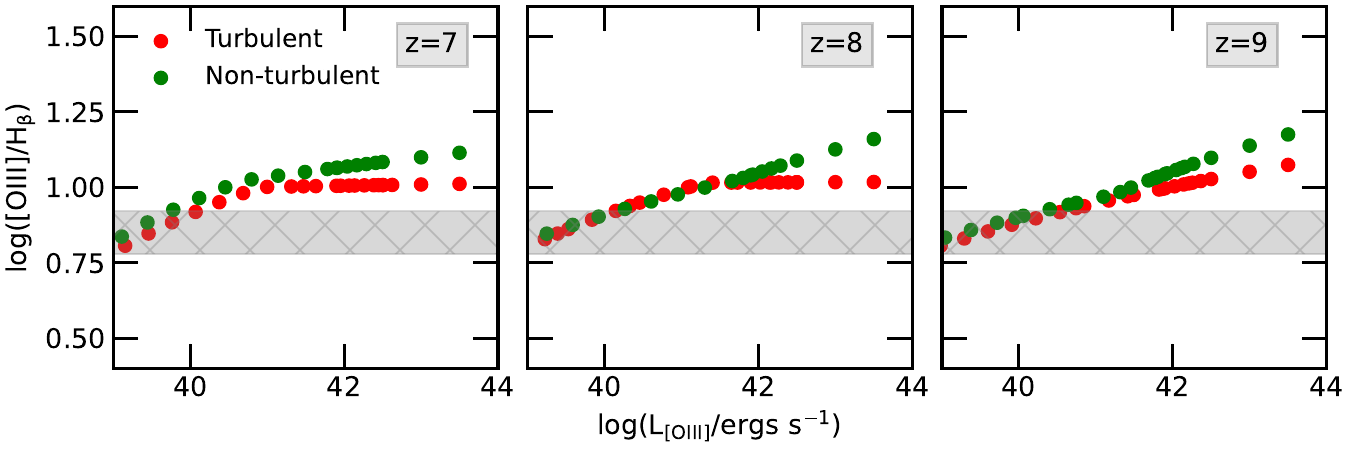}
    \caption{Flux ratio $\log{\OIII}/H_{\rm \beta}$ predictions plotted against $\log(L_{\OIII})$ for three different redshifts based on the \MV\ 5.2v photoionziation modelling code. The red and green dots correspond to turbulent and non-turbulent cases. The horizontal grey band correspond to measured ratio for the full sample stack of FRESCO $\OIII$ emitters.}
    \label{fig:line_ratio}
\end{figure*}

\begin{figure}
    \includegraphics[scale=0.56]{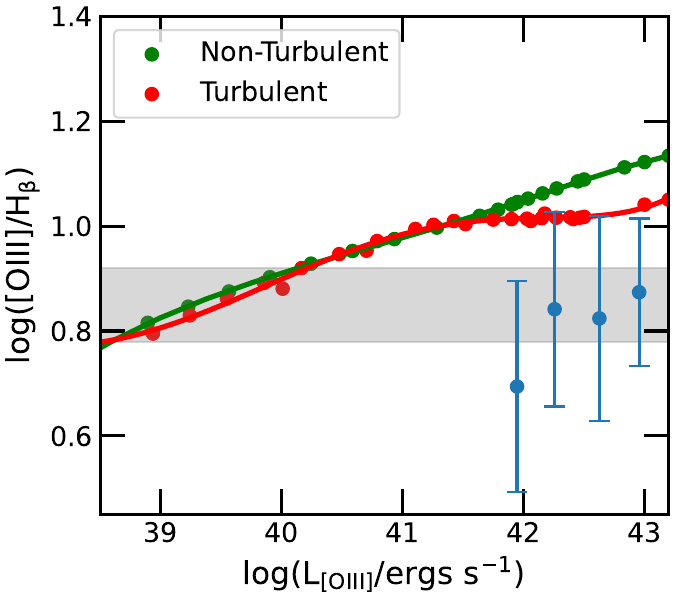}
    \caption{The weighted average (for $z =7,8,9$) of flux ratio, $\log{\OIII}/H_{\rm \beta}$ against $\log(L_{\OIII})$ based on the \MV\ 5.2v photoionziation modelling code. The red and green dots and best fit curve correspond to turbulent and non-turbulent cases respectively. Blue circles show the flux ratio of FRESCO \OIII\ emitters with the corresponding error bars.}
    \label{fig:weighted_line_ratio}
\end{figure}

\FloatBarrier
\section{Summary}
\label{sec:summary}
 
 We have developed an analytical model for the physical and observed properties of emission line galaxies spanning a redshift range from $z\sim 5-9$. The model is based on \citep{Stu2012} and calibrated against the SFRD function at $z \sim 6$ and observed metallicity at $z \sim 8$. We extend the model to calculate ISM density, pressure and ionization parameter and compute emission line luminosity  $\OIII \lambda 5007$ by using the model of \citep{Lidz20} and its relationship with UV luminosity and Star Formation Rate. We also incorporate the effect of ISM turbulence on ISM density. To validate our model predictions, we compare the relationship between UV and $\OIII \lambda 5007$ line luminosities finding the relation and estimated scatter from the buildup of metallicity to be in good agreement with observations from FRESCO. We present the model LF of $\OIII$ emitters, and also find good agreement with observations. \\

We also investigate the variation of flux ratio $\log{\OIII}/H_{\rm \beta}$ based on the $\MV$ v5.2.1 photoionization modelling code and use our calibrated model to make predictions for the ${\OIII} /H_{\rm \beta}$ flux ratios in high-z emission line galaxies. We find that the turbulent galactic discs are predicted to have a smaller ionization parameter and to have a smaller $\log{\OIII}/H_{\rm \beta}$ due to the less dense ISM. The predicted flux ratio prefers a turbulent ISM when compared with observations. Thus, emission line ratios may provide a probe of the high-z turbulent ISM. 

 \section{Acknowledgement}
This research was supported by ARC Centre of Excellence for All Sky Astrophysics (ASTRO 3D), through project \#CE170100013. PA thanks for insightful discussions and David Nicholls for providing insightful information on $\MV$ v5.2.1. We thank Pascal Oesch for kindly sharing the FRESCO catalog of {\OIII} emitters.

\section{Data availability}
The modelled data underlying this article will be shared upon reasonable request to the corresponding authors.


\bibliographystyle{mnras}
\bibliography{ref} 

\bsp	
\label{lastpage}
\end{document}